
\input harvmac
\baselineskip=12pt
\noblackbox

\def\ie{{\it i.e.}}
\def\nl{\hfil\break}

\def\npb{{ \sl Nucl. Phys. }}

\def\prd{{ \sl Phys. Rev. }}
\def\prl{{ \sl Phys. Rev. Lett. }}
\def\plb{{ \sl Phys. Lett. }}
\def\rmp{{ \sl Rev. Mod. Phys. }}
\def\IR{\relax{\rm I\kern-.18em R}}
\def\undertext#1{\vtop{\hbox{#1}\kern 1pt \hrule}}
\def\half{{1\over2}}
\def\c#1{{\cal{#1}}}
\def\dirac{\hbox{$\partial$\kern-0.5em\raise0.3ex\hbox{/}}}
\def\dslash{\hbox{$\partial$\kern-0.5em\raise0.3ex\hbox{/}}}
\def\pslash{\hbox{{\it p}\kern-0.5em\raise-0.3ex\hbox{/}}}
\def\gsim{\mathrel{\raise.3ex\hbox{$>$\kern-.75em\lower1ex\hbox{$\sim$}}}}
\def\lesssim{\mathrel{\raise.3ex\hbox{$<$\kern-.75em\lower1ex\hbox{$\sim$}}}}
\def\d{\partial}
\def\thw{\theta_W}
\def\nf{N_F}
\def\sb{s^{\scriptscriptstyle H}_{\scriptscriptstyle bare}}
\Title{\vbox{\baselineskip12pt\hbox{UCLA/92/TEP/23}\hbox{%
OHSTPY-HEP-T-92-008}}}
{ Non--decoupling, triviality and the $\rho$ parameter}
\centerline{Kenichiro Aoki\footnote{$^1$}
{Work supported by the National Science Foundation grant
NSF--89--15286.\nl internet:{\tt~aoki@physics.ucla.edu}}
and Santiago Peris\footnote{$^2$}{Work supported
by the Department of Energy Grant DOE/ER/01545-578.
Address after August 1992, Grup de Fisica Teorica, Universitat Autonoma de
Barcelona, E-08193 Bellaterra, Barcelona, Spain.
\nl internet:{\tt~peris@ohstpy.mps.ohio-state.edu}}}
\bigskip\centerline{\it ${}^1$Department of Physics}
\centerline{\it University of California at Los Angeles}
\centerline{\it Los Angeles, CA  90024{\rm --}1547}
\bigskip\centerline{\it ${}^2$Department of Physics}
\centerline{\it The Ohio State University}
\centerline{\it Columbus, OH  43210{\rm --}1106}
\vskip .3in
\centerline{\bf Abstract}
The dependence of the $\rho$ parameter on the mass of the Higgs scalar
and the top quark is computed non--perturbatively using the $1/N_F$ expansion
in the standard model.
We find an explicit expression for the $\rho$  parameter that
{\sl {requires}} the presence of a physical cutoff. This should come as no
surprise since the theory is presumably
trivial. By taking this cutoff into account,
we find that the $\rho$ parameter can take values only within a limited
range and has finite ambiguities that are suppressed by
inverse powers of the cutoff scale, the so called ``scaling--violations".
We find that large deviations
from the perturbative results are possible, but only when the cutoff
effects are also large.

\Date{6/92}
\newsec{Introduction}
In the standard model\ref\WSG{S.~Weinberg, \prl{\bf 19} (1967) 1264\nl
A.~Salam, in {\sl``Elementary particle theory: Relativistic
groups and analyticity",} Nobel Symp. No. 8, N.~Svartholm ed.,
Almqvist and Wiksell, Stockholm (1969) \nl
S.L.~Glashow, \npb{22} (1961) 579 }, all
particles acquire masses through their interaction with the vacuum. In
perturbation theory, these
masses are proportional to the corresponding coupling constant times the
vacuum expectation value of the scalar field.
For a fixed vacuum expectation value, the masses
are larger only if the coupling constants are stronger. In this situation,
the decoupling theorem does not apply\ref\DECOUPLING{T.~Appelquist,
J.~Carazzone,  \prd{\bf11} (1975) 2856 \nl J.C.~Collins,
{\sl ``Renormalization"}, Cambridge University Press (1984), page 223},
and there exists the interesting possibility of having physical effects at
low energy that do not vanish as a particle becomes heavier,
called
{\sl {non--decoupling}}  effects.
The most celebrated example of such an effect is the so called $\rho$
parameter\ref\VELTMAN{M.~Veltman, \npb{\bf B123} (1977) 89}%
\ref\RADIATIVE{For
a review on radiative correction in the standard model, see for instance,
K.~Aoki, Z.~Hioki, R.~Kawabe, M.~Konuma, T.~Muta, {\sl Suppl.
Progr. Theor. Phys.,} {\bf 73} (1982) 1.\nl
For a recent comparison with experiments, see for instance,
P.~Langacker, M.~Luo, A.~Mann, \rmp{\bf64} (1992) 87},
which not only does not vanish, but
grows with the mass of the top quark or the Higgs boson.
The $\rho$ parameter measures the relative
strength of the neutral current  to the
charged current interactions at low energy,
as measured, for instance, in
neutrino scattering experiments.

The top quark and the Higgs scalar are the only particles in the standard model
that are yet to be found.
At one loop,
the contribution to the $\rho$ parameter from the top grows quadratically
with its mass as
\eqn\rhotone{\rho\Bigr|_{top}\approx1+{3\over (4\pi)^2} \sqrt 2 G_Fm_t^2 }
where $m_t$ is the top mass and $G_F$ is the Fermi constant. However in the
case of the Higgs, the contribution has only a logarithmic dependence
on its mass in accordance with
the screening theorem\ref\SCREENING{M.~Veltman,
{\sl Acta Phys. Pol.}{ \bf B8}
(1977); \plb{\bf 70B} (1977) 253\nl M.B.~Einhorn, J.~Wudka,
\prd{\bf 39D} (1989) 39}
and reads
\eqn\rhohone{\rho\Bigr|_{Higgs}
\approx 1- {3\over 4} {g'^2\over (4\pi)^2}\log ({M_H^2\over M_W^2})}
where $g'$ is the U$(1)_Y$ weak hypercharge gauge coupling,
$M_H$ is the Higgs mass and $M_W$ is the $W$ mass.

As $m_t$ and $M_H$ grow, so do their respective contributions to $\rho$.
Since $\rho$ is measured to be unity to roughly a percent,
 this in principle can place bounds on $m_t$ and $M_H$. In practice,
it is only the top contribution that grows fast enough to violate
the experimental constraint within the range of applicability
of perturbation theory, under which \rhotone\  was derived. This gives the
ubiquitous bound of $m_t\lesssim 200 GeV$\ref\RPP{Review of
particle properties, \prd{\bf 45D} (1992) S1}
quoted in the literature. The Higgs contribution grows extremely
slowly and by the time it reaches the experimental limit, the Higgs mass
is far too large to be in the perturbative regime, so that \rhohone\  does not
give any useful bound on $M_H$. The fact that $\rho$ grows with $m_t$ and
$M_H$ is a manifestation of non--decoupling.

An a priori unrelated feature of the standard model
is the fact that there is evidence
that the non--gauge sector of the theory is ``trivial". This, very succinctly,
means that the theory has a built--in cutoff and the Higgs%
\ref\ON{K.G.~Wilson, \prd{\bf D7} 2911\nl
L.~Dolan, R.~Jackiw, \prd{\bf D9} (1974) 3320\nl
H.J.~Schnitzer, \prd{\bf D10} (1974) 1800,2042\nl
S.~Coleman, R.~Jackiw, H.D.~Politzer, \prd{\bf D10} (1974) 2491\nl
  L.F.~Abbott, J.S.~Kang, H.J.~Schnitzer, \prd{\bf D13} (1976) 2212\nl
W.A.~Bardeen, M.~Moshe, \prd{\bf D28} (1983) 1372, and references therein}%
\ref\LUSCHER{M.~Luscher, P.~Weisz, \npb{\bf B290} (1987) 25,
{\bf B295} (1988) 65, \plb{\bf 212B} (1988) 472}%
\ref\NUM{
A.~Hasenfratz, K.~Jansen, C.B.~Lang, T.~Neuhaus, H.~Yoneyama, \plb{\bf 199B}
(1988) 531\nl
J.~Kuti, L.~Lin, Y.~Shen, \prl{\bf61} (1988) 678\nl
J.~Jers\'ak, in {\sl``Eloisatron Project Workshop",}  (1988)}%
\ref\EINHORN{M.B.~Einhorn, \npb{\bf B246} (1984) 75}%
and the top%
\ref\EG{M.B.~Einhorn, G.~Goldberg, \prl{\bf57 }(1986) 2115}%
\ref\KA{K. Aoki, \prd{\bf D44} (1991) 1547}%
\ref\FLAT{J.~Shigemitsu, \plb{\bf226B} (1989) 364\nl
I-H.~Lee, J.~Shigemitsu, R.~Shrock, \npb{\bf B330} (1990) 225,
\npb{\bf B335} (1990) 265\nl
W.~Bock, A.K.~De, C.~Frick, K.~Jansen, T.~Trappenburg,
\npb{\bf B371}~(1992)~683\nl
S.~Aoki, J.~Shigemitsu, J.~Sloan, \npb{\bf B372} (1992) 361\nl
Proceedings of the International Symposium on Lattice Field Theory,
KEK (1991)\nl and references therein.}
masses have theoretical upper bounds set by requiring the internal
consistency of the theory.

In this paper we would like to study these non--decoupling effects
taking the $\rho$ parameter as an example, and
address the question of what happens to the top and the
Higgs contributions to the $\rho$ parameter when the couplings are so
large that  perturbation theory is not valid. Do they still grow with the
mass of the particle, or do they saturate somehow?
How does the presence of a physical cutoff affect these contributions? It
should become clear
that our results are applicable to
more general non--decoupling effects than just the contributions to $\rho$.

To try to answer these
questions we shall use the $1/N_F$ expansion where $N_F$ is the number of
components of the scalar field. (In the standard model
$N_F$ is two.)
The standard model needs to be generalized accordingly
as we shall explain below; the scalar
sector is none other than the classic O$(2N_F)$ model of \ON\EINHORN\
and the fermion sector was generalized in \KA.

The $1/N_F$ expansion entails a systematic truncation of the
Schwinger--Dyson equations and is
non--perturbative in the coupling constant.
It resums an infinite set of diagrams.
We shall see that a consequence of this resummation is a pole in the
loop momentum that develops in the
expression for the $\rho$ parameter. Because
of this pole, were we to integrate the loop momentum
from zero to infinity --- as we usually do ---
the expression would diverge.
This pole necessitates
the presence of a built--in cutoff, only up to which the integration is
to be performed.
We identify this momentum scale with the cutoff scale  obtained from  the
triviality arguments.
This further clarifies the crucial role played by
the cutoff in the theory, bringing up an intriguing connection between
non--decoupling and triviality.

Of course, once a cutoff is present there will be
effects that depend on the fine details of the actual cutoff procedure,
often referred to as  ``scaling violations".
These effects are completely negligible for all practical purposes within
the perturbative regime, where the cutoff is very large, since they are
suppressed by inverse powers of the cutoff. In the non--perturbative
regime, however, the cutoff
scale is not much higher than the physical momentum
scale under consideration and these scaling violations are not suppressed
at all. It is only in this situation
that we find large deviations from the perturbative results.

Also, the $1/\nf$ expansion for the $\rho$ parameter
includes graphs that intrinsically have arbitrary number of loops
in the resummation. Since the two loop computations have been
performed in the standard model both for the Higgs%
\ref\BV{J.~van~der~Bij, M.~Veltman, \npb{\bf B231}(1984) 205}
and for the top%
\ref\BH{J.J.~van der Bij, F.~Hoogeveen, \npb{\bf B283}
(1987) 477\nl M.~Consoli, W.~Hollik, F.~Jegerlehner,
\plb{\bf B227} (1989) 167}, our results provide a non--trivial
comparison of the results of the $1/\nf$ expansion
against those of perturbation theory.
We find that our simple expressions for the $\rho$ parameter
contain the essential features of the perturbative results.
Furthermore, by comparing the perturbative results to those derived
using the  $1/\nf$ expansion, the region where perturbation
theory is valid becomes clear.

There are already calculations of the $\rho$ parameter utilizing a
$1/N$--type expansion in the literature for the contributions of the Higgs%
\ref\CHO{P.~Cho, \plb{\bf 240B} (1990) 407},
and the top quark\ref\PERIS{S.~Peris, \plb{\bf 251B} (1990) 603}.
However, the role of the physical
cutoff in radiative corrections with
perturbative non--decoupling is, to the best of our knowledge,
a novel feature of the present work. The presence of a physical cutoff
was apparently not considered in \CHO; and in \PERIS, $\rho$ was
computed only to leading order in $1/N_{color}$
wherein the cutoff does not appear.

Admittedly, $N_F$=$2$ is not $N_F$=$\infty$. However, we think that our
results should be reliable, at least qualitatively.
The existence of a physical cutoff in the theory  is the most crucial
ingredient of our calculation and this is a quite firm result that
has been established
both using the lattice\LUSCHER\NUM\FLAT
and large--$N$ methods\ON\EINHORN\EG\KA.
The need to cut off the loop momenta
and the appearance  of scaling violations are natural consequences of
having a physical cutoff in the theory.
\newsec{The dynamics of the Nambu--Goldstone bosons and the
$\rho$ parameter}

At very low energies the standard model can be described by an effective
Lagrangian of the form
\def\gf{G_F}
\def\jcp{J_\mu^{+}}
\def\jcm{J_\mu^{-}}
\def\jnc{J_\mu^{0}}
\def\mw{M_W}
\def\mz{M_Z}
\eqn\ljj{\eqalign{-\c L_{JJ} &= {g^2\over 8\mw^2}\jcp{\jcm}
   +\half{g^2\over 8\mz^2 \cos^2\theta_W}\jnc\jnc\cr
 &={\gf\over\sqrt2}\left(\jcp{\jcm}+\half\rho\jnc\jnc\right)\quad.\cr}}
Here $J^{\pm}\equiv J_1\pm i J_2$, $J_0\equiv J_3-\sin^2\theta _W J_{em}$
where $J_i, (i=1,2,3)$ and $J_{em}$ denote
the isospin and electromagnetic currents respectively.
$\sin^2 \theta_W$ is the electroweak mixing angle
and $G_F$ is  the Fermi constant.
(Here and below, we use the spacelike metric $(-1,+1,+1,+1)$.)
Clearly from \ljj\ one finds the
following expression for the $\rho$ parameter
\eqn\rhodef{\rho=\left.{\mw^2\over\mz^2\cos^2\thw}\right|_{\rm zero\
momentum}\quad.}
At tree level, $\rho$ equals unity due to a global symmetry, the
so--called custodial
SU$(2)$ symmetry
under which $W^{\pm}$ and $Z/\cos \theta_W$ transform like a triplet
\ref\CUSTODIAL{P.~Sikivie, L.~Susskind, M.~Voloshin, V.~Zakharov,
 \npb{\bf B173} (1980) 189}.
Beyond the tree level one finds
\eqn\rhovacpol{\rho\approx 1- {\Pi_W\over M_W^2}+{\Pi_Z\over M_Z^2}}
where $\Pi_{W,Z}$ are the vacuum polarization functions of the $W,Z$ bosons,
and the approximation $\Pi_{W,Z}/M^2_{W,Z}\ll1$ has been used.

As discussed in the introduction, we would like to calculate
in the $1/N_F$ expansion
how $\rho$ is affected by a heavy top and a
heavy Higgs. Therefore, the non--perturbative effects one wants
to compute are caused by the Yukawa coupling, $y$, and the  quartic scalar
self-coupling, $\lambda$, and have nothing to do with the gauge sector of
the theory which only plays the rather passive role of transmitting these
effects into the $\rho$ parameter where we can measure them.
In fact, the SU$(2)_L$ and U$(1)_Y$
gauge couplings, $g$ and $g'$, are
small and ordinary perturbation theory in these couplings should be an
excellent approximation.

In the $1/N_F$ expansion, one always has the arbitrariness of how to
generalize the initial Lagrangian for arbitrary $N_F$. The above discussion
suggests that the obvious generalization to a SU$(N_F)_L$ (or Sp$(N_F)_L$ as in
Ref. \CHO)
gauge theory is an avoidable complication in this case. Our approach,
following \ref\LYTEL{R.S.~Lytel, \prd{\bf 22D} (1980) 505
}, will be to define the $\rho$ parameter in terms of the
Nambu--Goldstone fields,
making contact with what is actually  measured in the
standard model by first taking the limit $N_F\to 2$ and then gauging the
SU$(2)_L$ group at the very end. In this way the SU$(2)_L$ gauge bosons act
solely as external fields.

Let us briefly summarize then how one can compute the $\rho$ parameter in
terms of the Nambu--Goldstone bosons. The relevant part of the Lagrangian
is the kinetic term for these bosons and, when radiative corrections are
taken into account, reads
\def\zzero{Z_{\chi^0}}
\def\zplus{Z_{\chi^+}}
\eqn\ngb{-\c L_\chi=
\zplus\left|\d_\mu\chi^+-{gv\over2}W_\mu^+\right|^2
+\half\zzero\left(\d_\mu\chi^0-{gv\over2\cos\thw}Z_\mu\right)^2
+\hbox{ other terms}}
where the above combination is determined by the
SU$(2)_L\times$U(1)$_Y$ gauge invariance.
The vacuum expectation value of the Higgs field $v$
in general includes the  shifts due to quantum corrections
and is known to be about $246\,GeV$ in the standard model.
The ``effective'' masses for the $W$ and $Z$ are
\eqn\masswz{-\c L_{mass}=\zplus\left({gv\over2}\right)^2W_\mu^+W_\mu^-
   +\half\zzero\left({gv\over2\cos\thw}\right)^2Z_\mu^2\quad.}

These are the masses that appear in \ljj\ and therefore one can express the
$\rho$ parameter as
\eqn\rhodef{\rho=\left.{\zplus\over\zzero}\right|_{\rm zero\  momentum}
\simeq 1 +(\zplus - \zzero) \qquad \  for \qquad (Z_{\chi^{+,0}}-1) \ll 1}
in terms of the wave function renormalization constants of the Nambu--Goldstone
fields defined at zero momentum. We would like to emphasize that the first
equality in Eq. \rhodef\ is exact up to contributions $\c O(g^2,g'^2)$. We
would also like to point out here the advantage of the approach using
Nambu--Goldstone bosons in the
case of a possible future lattice calculation of
$\rho$, doing away with the non--Abelian gauge bosons altogether.
To leading order in the SU$(2)_L$ gauge
coupling $g$, one does not need to gauge SU$(2)_L$ at all.\foot{%
We cannot do without the U$(1)_Y$ gauge coupling $g'$, since this
is responsible for the custodial symmetry breaking in the case of the
Higgs.}

Of course, if one wished to do so, one could
also obtain $\rho$ by calculating the vacuum
polarization of the gauge bosons as in Eq. \rhovacpol\ since, after all,
both calculations are related by
gauge invariance. We  shall also do so where appropriate and, by comparing
with the calculation with the Nambu--Goldstone bosons,
we shall see  some interesting effects in
the non-perturbative regime due to the presence of the physical cutoff.
\newsec{The Higgs contribution to the $\rho$ parameter}
\subsec{The Higgs sector in the large--$\nf$ limit}
We first summarize the aspects of the Higgs sector
in the large--$\nf$ limit necessary
for our purposes, following  \ON\EINHORN.
The Lagrangian for the scalar sector of the standard model
with no gauge coupling is
\eqn\higgs{-\c L_\phi=\d_{\mu}\phi^\dagger\d_{\mu}\phi
+\lambda\left(\phi^\dagger\phi-v^2/2\right)^2\quad.}
Here, $\phi$ is in  a $\nf$ dimensional irreducible representation
of SU$(\nf)_L$.
Without any loss of generality, the scalar field develops
a vacuum expectation value $\langle\phi\rangle=(v/\sqrt2\ 0\ 0\ldots0)^T$.
This vacuum expectation value breaks the symmetry of the model
from O$(2N_F)$ to O$(2N_F-1)$. This O$(2N_F-1)$ unbroken symmetry rotates
all the Nambu--Goldstone bosons among themselves.
Therefore it plays the role of
the custodial symmetry in that $\rho$, as defined for instance in Eq.
\rhodef, is necessarily unity. We  define $\chi^0$ and $\chi^-$
as $\phi=((v+H+i\chi^0)/\sqrt 2, i\chi^-,\ldots)^T$ where $\phi$ is a vector
with $N_F$ components.
The spectrum consists of one
massive real scalar  and $2\nf-1$ Nambu--Goldstone bosons.
The mass of the scalar, $H$, is $\sqrt{2\lambda }v$ at tree level.

The large--$\nf$ limit is taken by keeping $\lambda\nf$ and $v^2/\nf$
fixed while taking $\nf$ to infinity.
The quantum corrections that contribute to leading order
in the large--$\nf$ limit
are the massless Nambu--Goldstone bubble graphs.
The vacuum expectation value is not corrected to this order.
\def\lb{\lambda_{\scriptscriptstyle bare}}
\def\lren{\lambda(s_0)}
The bare coupling constant, $\lb$, needs to be renormalized and we
define the renormalized coupling, $\lren$,  in the following manner
\eqn\higgscoupling{\lren={\lb\over1-{\lb\nf/(8\pi^2)}\ln{s_0/\sb}}\quad.}
Here, $\sb$ denotes a regulator dependent quantity whose explicit
expression is not needed in the subsequent analysis.
The renormalized coupling constant, as defined above, has
the meaning of the strength of scattering at the momentum (squared)
scale $s_0$.
The renormalization scale, $s_0$, is arbitrary.
Another definition of the renormalized coupling, used
in most lattice computations is through
the physical Higgs mass in units of the
vacuum expectation value.
In this case, the relation between the bare and the renormalized
coupling is not as explicit as the definition we use \higgscoupling.

The propagators for the Nambu--Goldstone bosons are not
modified to leading order in $1/\nf$.
However, the propagator of the Higgs field, $H$, receives the bubble
contributions as in \fig\figbubble{The leading order
one--particle irreducible quantum contributions in the $1/\nf$ expansion
to the Higgs propagator.}
and may be expressed in terms of the renormalized parameters as
\def\sehiggs{\Sigma_H}
\eqn\higgsprop{
D_H(p^2)=\left[p^2+\sehiggs(p^2) \right]^{-1},\qquad\hbox{where }\quad
\sehiggs(p^2)\equiv{2\lren v^2\over1-{\lren\nf/(8\pi^2)}\ln{p^2/s_0}}\quad.}
Notice that $\Sigma_H(p^2)$, and as a consequence $D_H(p^2)$, are
$s_0$-independent.
The quantum effects, roughly speaking, change the mass into a
momentum dependent mass.
The physical mass and width of the Higgs particle are determined
by the location of the pole of the above propagator in the complex
plane as $p^2=-\left(M_H-i \Gamma_H/2\right)^2$.

\def\strivh{s_{triv}^H}
At the momentum scale $\strivh\equiv s_0\exp(8\pi^2/(\lren\nf))$,
cross sections diverge. We identify this scale as the physical cutoff
scale in the theory, usually called the ``triviality scale".
This cutoff scale is a physical quantity that is independent
of the renormalization scheme chosen.
We note that the cutoff scale has a typical non--perturbative
dependence on the coupling.
As the renormalized coupling constant 
increases, at some point
the model becomes physically unacceptable for at least two reasons;
the cross sections at the scale of the mass of the Higgs diverge
and the cutoff scale of the theory becomes as small as the mass.
This leads to a bound on the Higgs mass of about
$1\,TeV$\LUSCHER\NUM\EINHORN.
The mass, width and the triviality scale are plotted against the renormalized
coupling in \fig\massplot{The mass, width,
and the triviality scale in units
of the vacuum expectation value, $v$,
 vs. the renormalized coupling $\lren$.
$y^2/4$ in the label for the $x$--axis
refers to the Yukawa coupling which appears
in the next section.}.
We also plot $0.1\,\sqrt{\strivh}$ for convenience.
In the plot, the renormalization scale $s_0$
was chosen to be $|p^2|=M_H^2+\Gamma_H^2/4$ where $p^2$ is the location of the
pole of the propagator \higgsprop\ in the complex plane.
In this case, the cutoff scale $\strivh$
becomes equal to  to the mass scale
$s_0=M_H^2+\Gamma_H^2/4$ and $M_H$ becomes equal
to $\Gamma_H/2$ when $\lambda(s_0)$ is infinity.
\subsec{The corrections to the $\rho$ parameter}
When
the  O$(2N_F-1)$ custodial symmetry is unbroken, $\rho=1$.
The interactions in the scalar potential of the standard model do
not break the custodial symmetry. The weak hypercharge, however, does break
this custodial symmetry and therefore one should expect corrections
to the $\rho$ parameter due to the scalars
to be of order $g'^2$, where $g'$ is the U$(1)_Y$ weak hypercharge gauge
coupling.
Therefore, we need to gauge U$(1)_Y$ to compute the $\rho$ parameter.
The Lagrangian is
\eqn\lphib{-\c L_{\phi B}={1\over4}B_{\mu\nu}^2
  +{1\over2a}\left(\d_\mu B_\mu\right)^2+|D_\mu\phi|^2
  +\lambda\left(\phi^\dagger\phi-v^2/2\right)^2}
where
\eqn\bdef{B_{\mu\nu}\equiv \d_\mu B_\nu-\d_\nu B_\mu,\qquad
D_\mu\phi\equiv(\d_\mu+ig'/2 B_\mu)\phi\quad.}
We let the gauge parameter $a$ go to zero and enforce the
transverse polarization, $\d\cdot B=0$, which eliminates the mixing
term between $\chi^0$ and $B_\mu$.
\def\mb{M_B}
In this gauge, the gauge boson $B$ has mass $\mb\equiv g'v/2$ and
the Nambu--Goldstone bosons are massless.

\def\piplus{\Pi_{\chi^+}^H}
\def\pizero{\Pi_{\chi^0}^H}
\def\piplusp{\Pi_{\chi^+}^{H\prime}}
\def\pizerop{\Pi_{\chi^0}^{H\prime}}
The leading order corrections to the Nambu--Goldstone propagators
arise from the graphs in \fig\fighgraph{The leading order
contributions to the Nambu--Goldstone propagators.
The blob denotes the full propagator.}
which we call $\piplus,\pizero$.
As usual, one may check that the class of graphs in
\fig\fighseagull{The seagull and Higgs tadpole graph contributions
to the propagators of Nambu--Goldstone bosons that cancel.}
that contribute to the mass of the Nambu--Goldstone bosons cancel.
Also there are contributions to the Nambu--Goldstone propagators
of $\c O(1/\nf)$
that do not depend on the U$(1)_Y$ coupling $g'$ and hence do
not contribute to the $\rho$ parameter at this order.
The Nambu--Goldstone propagators become
\def\dplus{D_{\chi^+}}
\def\dzero{D_{\chi^0}}
\eqn\ngprop{\dplus(p^2)=\left(p^2-\piplus(p^2)\right)^{-1},\qquad
\dzero(p^2)=\left(p^2-\pizero(p^2)\right)^{-1}}
so that
\eqn\zh{\zplus(p^2)=1-\piplusp(p^2),\qquad
\zzero(p^2)=1-\pizerop(p^2)}
where the prime denotes the derivative with respect to $p^2$.

A  simple computation secures the following
expressions for $\piplus,\pizero$.
\def\dk{\int\!\!{d^4\!k\over(2\pi)^4}\,}
\def\dkl{\int_{k^2<\Lambda^2}\!\!{d^4\!k\over(2\pi)^4}\,}
\def\ddk{\int\!\!{d^4\!k\over(2\pi)^4}\,}

\eqn\pieq{\eqalign{
\piplus(p^2)&=g'^2\dk{k^2p^2-(k\cdot p)^2\over ((k+p)^2+\mb^2)(k+p)^2k^2}\cr
\pizero(p^2)&=g'^2\dk
   {k^2p^2-(k\cdot p)^2\over((k+p)^2+\mb^2)(k+p)^2}D_H(k^2)\quad.\cr}}
$D_H$ is the Higgs propagator
in the large--$\nf$ limit in \higgsprop. Here and below, the integrals
have been continued to Euclidean space.
These expressions are logarithmically divergent in the
ultraviolet and they can be regularized, for instance,  by the
dimensional regularization.

Using \rhodef, we derive the $\rho$ parameter to
leading order in the U$(1)_Y$ gauge coupling as
\eqn\rhoh{\eqalign{\rho-1\biggr|_{Higgs}&={\zplus(0)\over\zzero(0)}-1
    = \pizerop(0)-\piplusp(0)+\c O(g'^2g^2,g'^4,g'^2/\nf)\cr
  &=-{3\over4}g'^2\dkl
   {\sehiggs(k^2)\over k^2(k^2+\mb^2)(k^2+\sehiggs(k^2))}\quad.\cr}}
The $\rho$ parameter is expressed in terms of renormalized
parameters and is independent of the renormalization scale $s_0$,
as it should be.
A crucial comment is in order; the above integral is
integrable both at $k^2=0$ and $k^2=\infty$. However it is {\it not}
integrable in the region in between.
There is a pole in the region $\strivh< k^2<\infty$. The reason for this
disease can be traced back directly to the existence of a pole in the
expression \higgsprop\ for the Higgs self--energy, which in turn is a
direct consequence of the pole in Eq. \higgscoupling, \ie\ of the Landau
pole.
As was discussed above, this region above the triviality scale is
not physically consistent and we shall cutoff the
integral at $\Lambda^2$, below the triviality scale $\strivh$.
It is important to note that it is not only physically
reasonable to cut off the integral
but it is {\sl necessary}.
The $\rho$ parameter in \rhoh\ defined with this
cutoff for the integral, is well defined and manifestly finite.
If the expression for the $\rho$ parameter is expanded with
respect to the coupling, $\lambda $, as in perturbation theory,
then one can integrate up to
infinity. The need to impose the cutoff does not appear to any order
in perturbation theory.

\def\mh{M_H}
Since the cutoff is now at a well defined finite
momentum scale, the physical parameters develop explicit
ambiguities that have to do with the details of the cut--off procedure
which are somewhat arbitrary.
These ambiguities
are of order $\c O(p^2/\Lambda^2)$ or $\c (\mh^2/\Lambda^2)$,
where $p^2$ is the momentum scale of the process
under consideration.
These suppressions are essentially guaranteed by the
renormalizability of the theory.
These ambiguities correspond to the ``scaling violations"
in the corresponding lattice theory.

\def\pwh{\Pi_W^H}
\def\pzh{\Pi_Z^H}
\def\mwt{M_{W,tree}}
\def\mzt{M_{Z,tree}}
The $\rho$ parameter may also be obtained computing the
pieces proportional to $\delta_{\mu\nu}$
in the self energy of the $W,Z$ gauge bosons in \fig\figwz{Leading
order self energy contributions
to the gauge bosons due to the Higgs.}, $\pwh,\pzh$.
(We use the unitary gauge.)
We obtain
\eqn\sewz{{\pwh(0)\over\mwt^2}=g^2\ddk {1+{k^2\over4\mwt^2}
  \over k^2+\mwt^2}D_H(k^2),\quad
{\pzh(0)\over\mzt^2}=g^2\ddk {1+{k^2\over4\mzt^2}
  \over k^2+\mzt^2}D_H(k^2)\quad.}
Here, we defined
$\mwt\equiv gv/2$, $\mzt\equiv gv/(2\cos\thw)$.
The additional seagull contributions of \fig\wzseagull{The seagull graph
contributions for the gauge bosons due to the Higgs.}
cancel exactly in their contributions to the $\rho$ parameter.
Then up to Higgs independent contributions, one finds
\eqn\rhohwz{\eqalign{\rho-1\biggr|_{Higgs}\!\!
      &={\pzh\over\mzt^2}-{\pwh\over\mwt^2}\biggr|_{Higgs}\cr
  &=-{3\over4}g'^2\dkl
   {\sehiggs(k^2)\over (k^2+\mwt^2)(k^2+\mzt^2)(k^2+\sehiggs(k^2))}\quad.\cr}}
This agrees with the expression obtained using the scalars in \rhoh\
up to terms that do not depend on the Higgs propagator and terms
that are of order $\c O(\mw^2/\mh^2,\mz^2/\mh^2)$. The limit of this
expression for
large $M_H$, which is what we are interested in, is the same as that of
Eq. \rhoh.

\def\mhp{M_{H,pert}}
It is instructive to compare the results with the perturbative results.
We may expand the expression for
the $\rho$ parameter obtained using the $1/\nf$ expansion in \rhoh\
in powers of $\lambda(\mhp^2)$ as
\eqn\rhohexp{\rho-1\bigr|_{Higgs}=- \left({g'\over4\pi}\right)^2
\left[{3\over4}\ln{\mhp^2\over\mb^2}
+ \left({\gf\mhp^2\over16\pi}\right)^2
+\c O(\lambda^3(\mhp^2))\right]}
where we introduced  $\mhp^2\equiv2\lambda(\mhp^2)v^2$.
$\mhp^2$ is the mass of the Higgs particle defined as the
pole of the propagator to one loop, which is
sufficient for the comparison with the two loop result.
We have expanded up to three loop order
and kept only
contributions that grow with the mass of the Higgs.
We may compare this to the expression up to two loops
given in \BV
\eqn\rhohpert{\rho-1\bigr|_{Higgs,2-loop}=-\left({g'\over4\pi}\right)^2
\left[{3\over4}\ln{\mhp^2\over\mw^2}
-5.37\times10^{-3}\gf\mhp^2+\c O(\lambda^2)\right]\quad.}
When comparing these two results,
we need to keep in mind that by definition, what we call ``the
Higgs contribution to the $\rho$ parameter" is well defined only up to
Higgs mass independent terms of order $g'^2$.
The one--loop term of course agrees, but  the order $\lambda$
term seen in the two--loop result \rhohpert\
is not present to this order in the $1/\nf$ expansion \rhohexp.
In the perturbative result of \BV, it was remarked that the coefficient
for this term has a strong cancellation and this is perhaps the
reason why it is not seen in our computation.
The results of the $1/\nf$ expansion are in agreement, as it should be,
with the screening theorem \SCREENING.

We plot the $\rho$ parameter obtained in the $1/\nf$ expansion
against the coupling $\lren$ and the Higgs mass in
\fig\rhohplot{The
Higgs contribution to the $\rho$ parameter in the $1/\nf$ expansion
plotted against the coupling, $\lambda(M_H^2+\Gamma^2/4)$.
``Cutoff" denotes the factor $\Lambda/\sqrt{\strivh}$.}
and \fig\rhohmplot{The
Higgs contribution to the $\rho$ parameter in the $1/\nf$ expansion
plotted against the mass in units of $v$.
``Max. mass" denotes the maximum mass of Higgs as seen from
the spectrum \massplot.}, respectively.
For illustrative purposes, we use two values for the cutoff,
$\Lambda/\sqrt{\strivh}=0.5$ and $\Lambda/\sqrt{\strivh}=0.1$.
For comparison, the one and two loop results are also shown.
We chose the same convention in \rhohplot\ for the
renormalization scale $s_0=M_H^2+\Gamma_H^2/4$ as
in \massplot.

The $1/\nf$ result for the $\rho$ parameter agrees very well
with the perturbative results when the coupling constant
is small and deviates from them for larger larger values of the
coupling.
At some point, the cutoff scale becomes as small as the
momentum scale of the mass, $M_H^2+\Gamma_H^2/4$,
at which point the results cease to make sense and the
$\rho$ parameter is plotted only up to this coupling.
The deviations from the perturbative results and the scaling
violations are clearly visible in the plot of $\rho-1$
against the coupling in \rhohplot.
However, from the plot against the mass in \rhohmplot,
we see that the large deviations from the perturbative
results and the scaling violations occur only in the
region where the mass is almost at its triviality bound.
The maximum of the
mass of the Higgs in the $1/\nf$ expansion  is $3.5\,v$.
(See \massplot.)
Typical numbers we obtain are the following:
The relative difference between $\rho-1$ computed with the cutoffs
for the integral $\Lambda/\sqrt{\strivh}
=0.1,0.5$ is 10\%\ at $M_H/v=3.3$. At this point, $\lren=6.9$,
$\sqrt{\strivh}/v=60$,
$\Gamma_H/M_H=0.59$ and the discrepancy from the two--loop result
is few in $10^{-4}$ which is of the order of the terms not enhanced
with respect to the Higgs mass which we did not compute.
When the results computed for the two cutoffs differ by
1\%, $M_H/v=2.9$, $\lren=4.6$,
$\sqrt{\strivh}/v=220$ and  $\Gamma_H/M_H=0.44$.
We notice that there is a small region in which the
$1/\nf$ result differs from the perturbative result
while the cutoff effects are small.
It will be interesting to see what happens when the
experimental precision reaches the level needed to
see the cutoff effects.

$1/\nf$ is not that small in the standard model and the rough
trend of the corrections may be deduced by comparing the $1/\nf$
results to the one loop
results as was discussed in \ref\LKS{L.~Lin,
J.~Kuti, Y.~Shen, in  the proceedings of {\sl``Lattice
Higgs Workshop"}, Tallahassee (1988)}.
The expression for the renormalized coupling constant,
$\lambda(s_0)$, should be contrasted to the one obtained
by integrating the one--loop renormalization group equation, in which the
only difference is that the factor $N_F$ in Eq. \higgscoupling\ is replaced
by $(N_F+4)$.
At the same time, in the contribution of the Nambu--Goldstone
bosons to the Higgs propagator \higgsprop,
the factor $\nf$ should be replaced by $(\nf-1/2)$ at one loop.
The effect of this discrepancy between the two factors of $\nf$
may be summarized by saying that for a given
Higgs mass, the cutoff is smaller
than the estimate given by the large--$\nf$ limit.
The most significant net effect is to make the physically
acceptable region smaller, and it does not change
the results qualitatively.

Our results seem to be in agreement with \CHO, where
cutoff effects were not discussed.
However, it is difficult to make a detailed comparison.
In the region of interest,
when the $\rho$  parameter differs from the perturbative
results substantially, the cutoff effects are important
and at some point the cutoff becomes as small as the Higgs mass.
\newsec{The top contribution to the $\rho$ parameter}
\subsec{The quark--Higgs sector in the large--$\nf$ limit}
The classical Lagrangian for the quark--Higgs sector
when a quark is massive is
\def\ql{q_{\scriptscriptstyle L}}
\def\bl{b_{\scriptscriptstyle L}}
\def\tl{t_{\scriptscriptstyle L}}
\def\tr{t_{\scriptscriptstyle R}}
\eqn\clphiq{-\c L_{\phi q}=-\c L_\phi+\overline\ql\dslash\ql
+\overline\tr\dslash\tr
+y\left(\overline\ql\phi\tr+\overline\tr\phi^\dagger\ql\right)\quad.}
$\ql$ and $\tr$ are irreducible representations of SU$(\nf)_L$ of
dimensions $\nf$ and $1$, respectively.
The model has a SU($\nf)_L\times$U$(1)_L\times$U$(1)_R$ symmetry
under the following transformations
\def\all{\alpha_L}
\def\alr{\alpha_R}
\eqn\qtransf{\ql\mapsto U_Le^{i\all}\ql,\quad
\tr\mapsto e^{i\alr}\tr,\quad\phi\mapsto U_Le^{i(\all-\alr)}\phi,
\quad U_L\in{\rm SU}(\nf)_L,\ \all,\alr\in\IR\quad.}
The vacuum expectation of $\phi$ breaks the symmetry to
SU$(\nf-1)_L\times$U$(1)_{L+R}$ and gives mass to the $t$--fermion.
We rewrite the $\ql$--field for convenience as $\ql\equiv(\tl \bl^1\
\bl^2\ldots\bl^{\nf-1})^T$.
The spectrum consists of one massive Dirac fermion, $t\equiv\tl+\tr$, and
$\nf-1$ Weyl fermions $\bl^i$.
Classically, the mass of $t$ is $yv/\sqrt2$.
When the Yukawa coupling $y$ vanishes, one
recovers the O$(2N_F)\to $O$(2N_F-1)$ symmetry breaking pattern of the pure
scalar theory. The Yukawa coupling breaks the custodial symmetry
O$(2N_F-1)$ and will yield corrections to the $\rho$ parameter.

The large--$\nf$ limit is taken by keeping $y^2\nf,v^2/\nf$ fixed
as we take $\nf$ to infinity.
In the large--$\nf$ limit, the only one--particle irreducible
graphs that contribute are the  corrections to the $t$--propagator
in \fig\figtprop{Leading order one--particle irreducible
contributions to the $t$ propagator
in the large--$\nf$ limit.}.
The full propagator for the $t$--field, in terms of the
bare quantities, is
\def\yren{y^2(s_0)}
\def\arb{A_{\scriptscriptstyle R,bare}}
\def\pl{P_L}
\def\pr{P_R}
\def\yb{y_{\scriptscriptstyle bare}}
\def\sbt{s_{\scriptscriptstyle bare}^{\scriptscriptstyle t}}
\eqn\tprop{S_t(p)=\left\{i\pslash\left[\arb(p^2)\pr
          +\pl\right]+\yb v\right\}^{-1},\qquad
\arb(p^2)\equiv1-{\yb^2\nf\over2(4\pi)^2}\ln{p^2\over\sbt}\quad.}
Here, $\pl,\pr$ are projection operators onto the left, right--handed
fields and $\sbt$ denotes a regulator dependent quantity whose explicit
expression we shall not need.
In particular, there are no corrections to this order to
the scalar sector of the model so that the two sectors
may be solved independently.
The coupling constant $\yb$ is renormalized similarly to the
scalar self coupling (cf. \higgscoupling)
\eqn\yrenorm{y^2(s_0)={\yb^2\over1-\yb^2\nf/(32\pi^2)\ln s_0/\sbt}}
where the renormalization scale $s_0$ is arbitrary.
The renormalized coupling corresponds to the scattering
strength at the momentum (squared) scale $s_0$.

\def\strivt{s_{triv}^t}
This coupling diverges at the scale $\strivt\equiv s_0\exp
\{32\pi^2/(y^2(s_0)\nf)\}$ and we identify this scale $\strivt$ as
the physical cutoff scale in the theory.
The mass, width and the cutoff scale may be determined numerically
and is identical to the Higgs case in \massplot\  provided
we make the replacement $\lambda(s_0)\mapsto \yren/4$.
(In the plot, the renormalization scale $s_0$ should be understood to be the
norm of the pole of the propagator in \tprop\ to agree with the conventions
of the Higgs case.)
For details on the solution of this model, we refer to \KA.
As in the Higgs case,
there is a bound on the top mass of around $1\,TeV$
in the $1/\nf$ expansion\KA.
When the custodial symmetry is not broken, the mass bound
was established in \EG.
Also, a considerable amount of numerical work exists
in the lattice approach that supports this point of view\FLAT,
although the results should be viewed as still being preliminary.
\subsec{The corrections to the $\rho$ parameter}
The contribution of the top to the $\rho$ parameter
is qualitatively different from that of the Higgs.
The Yukawa coupling of the fermions
breaks the custodial symmetry so that the correction to the
$\rho$ parameter is not proportional to the gauge coupling.
Furthermore, no analog of the screening theorem exists so
that the correction may become an order of magnitude larger
than the case of the Higgs.
The leading order corrections to the Nambu--Goldstone propagator
are of order $\c O(1/\nf)$ and arise from the
graphs in \fig\figngt{The leading order top
corrections to the Nambu--Goldstone propagators.
The blobs denote the full propagators.}.
These  graphs may be computed simply and are expressed
using the renormalized parameters as
\def\pitplus{\Pi_{\chi^+}^t}
\def\pitzero{\Pi_{\chi^0}^t}
\def\pitplusp{\Pi_{\chi^+}^{t\prime}}
\def\pitzerop{\Pi_{\chi^0}^{t\prime}}
\def\nc{N_c}
\def\ar{A_R}
\def\mt{m_t}
\eqn\ngtop{\eqalign{\pitplus(p^2)&=2\yren\nc\dk{k(k+p)\over
(k+p)^2\left[\ar(k^2)k^2+\hat \mt^2\right]}\cr
  \pitzero(p^2)&=2\yren\nc\dk{\ar((k+p)^2)k(k+p)+\hat \mt^2\over
\left[\ar(k^2)k^2+\hat \mt^2\right]
\left[\ar((k+p)^2)(k+p)^2+\hat \mt^2\right] }\cr}}
where $\nc$ is the number of colors, $\hat \mt^2\equiv\yren v^2/2$
and $\ar(s)\equiv1-\yren\nf/(32\pi^2)\ln s/s_0$.
The reader is cautioned that $\hat \mt$ is related, but not equal to,
the mass of the top.
At the same order, there is a tadpole
graph for the Higgs field $H$ as
in \fig\figtadpole{The tadpole contribution
due to the top for $H$ of order $1/\nf$.}
which leads to a shift in its vacuum expectation value.
The shift in the vacuum expectation value removes the mass term
from the Nambu--Goldstone fields.
(See, for instance, \ref\TAYLOR{J.C.~Taylor, {\sl ``Gauge theories
of weak interactions",} Cambridge University Press (1976)
page 124 and ff.}.)
Since this effect is momentum independent, it does not at all affect
our calculation of the momentum dependent pieces below.

Using \ngtop,
 the expression $\zplus=1-\pitplusp(0),\zzero=1-\pitzerop(0)$
and the equation \rhodef\ for the $\rho$ parameter, we derive
\eqn\rhotcorr{\rho-1\biggr|_{top}=\yren\nc\dkl{\hat \mt^2
\left(k^2\,\yren\nf/(32\pi^2)+\hat \mt^2\right)\over
k^2\left(\ar(k^2)k^2+\hat \mt^2\right)^3}\quad.}
As in the Higgs case, the integral needs to be cut off at a scale
$\Lambda^2\lesssim\strivt$. Then the above expression is well defined.

Just as in the Higgs case, we may compute the $\rho$ parameter
directly from the two point functions of the $W$ and $Z$ gauge  bosons.
The custodial symmetry breaking effects come both
from the Yukawa coupling and the U$(1)_Y$ weak hypercharge
coupling. Since we are only interested in the
leading effect which is not suppressed by the gauge
coupling, we may set $g'$ to zero in the calculation.
This simplifies the computation and
the two self energy graphs in \fig\figtwz{Self
energy graphs for the gauge bosons due to the top.}
yield the following expression for the $\rho$ parameter
\eqn\rhotwz{\rho-1\biggr|_{top}={\nc\over v^2}
\int_{k^2<\Lambda^2}\!\!{d^4\!k\over(2\pi)^4}\,
{\hat \mt^4\over k^2\left(\ar(k^2)k^2+\hat \mt^2\right)^2}\quad.}

Unlike the Higgs case, this
 expression differs from the one obtained using the Nambu--Goldstone
bosons in \rhotcorr.
Notice that, in general, one should not expect these two expressions to
agree since our naive implementation of the cutoff $\Lambda$ with a step
function breaks gauge invariance explicitly.
We do not know how to improve on this within
the present non--perturbative context.
If we use integration by parts and ignore the surface terms,
the two expressions agree. Since the surface terms are non--vanishing only
because of the existence of the finite cutoff, both forms are the same
to all orders within perturbation theory.
Once the cutoff effects are considered,
they agree only up to terms of order $\c O(\mt^2/\Lambda^2)$.
When the two formulas disagree by an appreciable amount,
we are in the region where the scaling violations also are substantial.

We expand the expression for the $\rho$ parameter
obtained using the $1/\nf$ expansion in \rhotcorr\ or \rhotwz\
in the coupling constant up to three loop order as
\def\mtp{m_{t,pert}}
\eqn\rhotexp{\rho-1\bigr|_{top}=\nc\left[x+\nf x^2 +
\nf^2\left(\half+{\pi^2\over6}\right)x^3+\c O(x^4)\right],\quad
\hbox{where \ \ }x\equiv{\gf\mtp^2\over8\sqrt2\pi^2}}
with $\mtp^2\equiv y^2(\mtp^2)v^2/2$ analogously to the Higgs case,
and compare with the result up to two loops\BH
\eqn\rhotpert{\rho-1\bigr|_{top,2-loop}=
\nc\left[x+(\nc+(19-2\pi^2))x^2+\c O(x^3)\right]\quad.}
The two expressions are qualitatively in agreement.
The order $x$ term agrees and the coefficient for the
order $x^2$ term has the same
sign while the magnitude differs by 13\%.

The $\rho$ parameter may be computed numerically using
the above expressions \rhotcorr\ (``NG") or \rhotwz\ (``gauge")
and is compared to the one and two loop results in
\fig\rhotplot{Contribution of the top to the $\rho$ parameter plotted
against the Yukawa coupling $y^2(s_0)$ in
comparison with the 1,2--loop results.
Cutoff denotes the factor $\Lambda/\sqrt{\strivt}$. ``NG" and ``gauge"
stands for the calculation using the Nambu--Goldstone bosons and
the gauge bosons respectively.}
and
\fig\rhotmplot{Contribution of the top to the $\rho$ parameter plotted
against the mass in units of $v$ in
comparison with the 1,2--loop results.
``Max. mass" denotes the maximum mass of top as seen from the spectrum
\massplot.}.
In the plot of $\rho-1$ against the coupling in \rhotplot,
the renormalization scale $s_0$ was chosen to be the
norm of the pole of the $t$ propagator \tprop, analogously to the
Higgs case.
The $\rho$ parameter obtained using the $1/\nf$ expansion
is plotted only in the region $m_t^2+\Gamma_t^2/4<\Lambda^2$.
We use two values for the cutoff, $\Lambda/\sqrt{\strivt}=0.5$ and
$\Lambda/\sqrt{\strivt}=0.1$.
Even though the contribution to the $\rho$ parameter is
an order of magnitude larger than that of the Higgs, the
gross features are similar.
When the coupling constant is small, all methods of computation
agree, as they should.
The effect of using different cutoffs, the deviation
from the two loop results and the discrepancy between  the
calculation using the scalars and the gauge bosons are clearly
visible in the region where the coupling constant is large.
The scaling violations are appreciable only when the
mass of the top is also close to its maximal value
in which case the triviality scale is also not much larger.
There is a small region around $m_t/v\sim2.3$ where the cutoff effects
are small and the deviation from the two loop result is substantial.
The $1/\nf$ expansion perhaps captures the essence of higher loop effects
and it would be interesting to compare with the three loop result if
such is computed. The interest in this point,
however, is only theoretical, since the $\rho$
parameter is already measured to be one within a percent.

As in the Higgs case, we list some typical numbers:
When the relative discrepancy
between the $\Lambda/\sqrt{\strivt}=0.1,0.5$ is
10\%, $m_t/v = 2.9$. At this point, $y^2(s_0)=20$, $\sqrt{\strivt}/v=160$
and $\Gamma_t/m_t=0.42$. The calculation using the scalars and the gauge
bosons differ by 4\%\ and the $1/\nf$ result differs from
the perturbative results by 30\%.
When the relative difference between the results for the two cutoff scales
$\Lambda/\sqrt{\strivt}=0.1,0.5$ is 1\%,
$m_t/v = 2.6$, $y^2(s_0)=14$, $\sqrt{\strivt}/v=670$
and $\Gamma_t/m_t=0.30$. The scalar and the gauge calculations differ
by 0.5\%, and both differ from the two loop result by 9\% at this point.
\newsec{Discussion}
There is evidence that the
 standard model without gauge couplings needs to be
defined with a physical cutoff at a finite momentum scale.
This is clearly displayed in the $1/\nf$ expansion of
the standard model.
In this work, we computed the contribution of the Higgs
and the top to the $\rho$ parameter in the standard model using
the $1/\nf$ expansion.
The $\rho$ parameter is not finite unless the cutoff
is imposed also on the intermediate physical states.
We could treat the two contributions separately since
the corresponding couplings do not influence each other
in the large--$\nf$ limit, apart from the fact that
the cutoff of the combined theory is the lower of the two
cutoffs.
In particular, the contribution to the $\rho$ parameter
from the Higgs and the top may just be summed.

The existence of the physical cutoff at a finite scale
leads to a qualitatively different behavior from
perturbation theory, in the region of strong coupling.
For one thing, it leads to ambiguities in the physical
predictions of the theory. These ambiguities are of
the order of
the mass scale over the cutoff scale.
But more interestingly,
the growth of the non--decoupling effect
with respect to the particle mass that one observes
in the perturbative regime is saturated.
Heuristically, one may understand this fact by looking,
for instance, at the expression for the $\rho$ parameter
derived from the vacuum polarization of the gauge bosons \rhotwz.
Making the crude approximation of letting $\ar(k^2)\rightarrow1$,
we may perform the integral explicitly to obtain
\eqn\heur{\rho-1\biggr|_{top}\approx{\nc\over2(4\pi)^2}
y^2(s_0){\Lambda^2\over\Lambda^2+y^2(s_0)v^2/2}\qquad.}
In the perturbative regime, for $y(s_0)$ very small and consequently
$\Lambda$ extremely large, one finds the well known growth
with the Yukawa coupling, $\delta\rho\sim y^2$.
However, for $y(s_0)$  very large, one notices that $y(s_0)v$ can
overcome $\Lambda$ (which actually tends to a constant value
for large $y(s_0)$) so that in the non--perturbative regime one
finds instead that $\delta\rho\sim\Lambda^2/v^2$, \ie\ it
is $y(s_0)$--independent.
Of course, the latter result is cutoff dependent in a most
obvious way so that cutoff effects will be of order one in
this regime. This is indeed what we see, for instance, in
\rhotplot\ and \rhotmplot.
The same approximation applied to $\delta\rho$ obtained from
the Nambu--Goldstone boson propagators \rhotcorr\ agrees
with the previous $\delta\rho$ for $y\ll1$ while it
is a factor of two larger when $y^2(s_0)v^2\gg\Lambda^2$, again
in good agreement with the plots.
Similar arguments applied to the Higgs case yield
\eqn\heurrho{\rho-1\biggr|_{Higgs}\approx-{3\over4}\left(g'\over4\pi\right)^2
\left[\ln{8\lambda(s_0)\over g'^2}-\ln{\Lambda^2+2\lren v^2\over\Lambda^2}
\right]\quad.}
We again see that for $\lren v^2\gg\Lambda^2$
the result is $\lren$--independent but cutoff dependent, while
for $\lren\ll1$ the result reduces to the one loop expression.

It should perhaps be emphasized that substantial scaling violations
can  occur for non--decoupling effects
even at low energy\foot{For example,
in the case of the $\rho$ parameter, the relevant  energy
scale is effectively zero.}
since they  can behave as the mass over the cutoff.
This is unlike the case of decoupling effects
where scaling violations are of the order of the momentum scale over the
cutoff and therefore vanish at low energies.

The non--decoupling effects may become very different
than the perturbative results if the only requirement is
that the mass of the particle be smaller than the cutoff scale.
We may further require that the cutoff effects be small so  that
the cutoff scale is much larger
than the mass scale.
The smaller we demand the cutoff effects to be, the deeper
we are pushed into the perturbative regime.
These properties were established by analyzing the $\rho$
parameter in detail. However,
we believe that
the qualitative features are shared by non--decoupling effects
in trivial theories in general.

In the present work, we employed the characterization of the
$\rho$ parameter entirely in terms of the Nambu--Goldstone bosons.
This is conceptually simpler and, we believe,
it will prove very useful in the event of a future lattice
calculation.

One of our goals was to determine whether the effects of
a heavy particle grow unboundedly with the mass of
the particle as in perturbation theory or, as we have found,
are tamed by non--perturbative effects.
This question is of paramount importance for the fundamental
question of defining the standard model on the lattice and
its associated problem of non--decoupling effects due
to the fermion doublers.
Non--decoupling effects due to fermion doublers in the lattice
formulation of the standard model have recently been discussed in
\ref\BD{M.~Dugan, L.~Randall, MIT preprint MIT--CTP--2050 (1991)\nl
T.~Banks, A.~Dabholkar, Rutgers preprint RU--92--09 (1992)
\nl S.~Aoki, Tsukuba University preprint, UTHEP--233 (1992)
\nl M.~Golterman, D.~Petcher, lecture given at Rome lattice conference (1992)
\nl and references therein.
}.
\bigskip\noindent{\it Acknowledgments:}
We would like to thank J.J.~van~der~Bij for an explanation of his results,
Marty Einhorn and  Volodya Miransky for comments on our manuscript and
Dallas Kennedy for conversations.
We would also like to thank Robert Shrock for comments and his
interest in this work.
We would  especially like to thank Maarten Golterman, Junko Shigemitsu
and John  Sloan for numerous discussions and comments on our paper.
\listrefs
\listfigs
\end